\documentclass[12pt,twoside,a4paper]{article}
\usepackage{qmc,psfig}

\newcommand\half{\textstyle{1\over2}}

\newcommand{\Sign}{\mbox{Sign}}
\newcommand{\cxu}{c_{x,\uparrow}}
\newcommand{\cxd}{c_{x,\downarrow}}
\newcommand{\nxu}{n_{x,\uparrow}}
\newcommand{\nxd}{n_{x,\downarrow}}
\newcommand{\cyu}{c_{y,\uparrow}}
\newcommand{\cyd}{c_{y,\downarrow}}
\newcommand{\nyu}{n_{y,\uparrow}}
\newcommand{\nyd}{n_{y,\downarrow}}

\begin{document}

\title{Novel Quantum Monte Carlo Algorithms for Fermions}
\author{
Shailesh Chandrasekharan}
\instit{
Department of Physics, Duke University,\\
Durham, NC 27704-0305, USA}
\gdef\theauthor{S.\ Chandrasekharan}
\gdef\thetitle{Novel Quantum Monte Carlo Algorithms for Fermions}
\maketitle

\begin{abstract}
Recent research shows that the partition function for
a class of models involving fermions  can be written as a statistical 
mechanics of clusters with positive definite weights. This
new representation of the model allows one to construct novel
algorithms. We illustrate this through models consisting of
fermions with and without spin. A Hubbard type model with both 
attractive and repulsive interactions becomes tractable 
using the new approach. Precision results in the two dimensional 
attractive model confirm a superfluid phase transition in the 
Kosterlitz-Thouless universality class.
\end{abstract}

\section{Introduction}

Fermion algorithms are known to be notoriously difficult. The 
main reason for this is that it is difficult to write the partition 
function of models involving fermions as a sum over configurations with 
positive definite weight. The most common approach is to integrate out the 
fermions in favor of a fermion determinant which leads to an effective 
bosonic partition function. Typically one obtains
\begin{equation}
Z \;=\; \int \;[d\phi]\;\exp\left(-S[\phi]\right)\;\mathrm{Det(M[\phi])}
\end{equation}
where the determinant is a non-local function of the bosonic fields 
$[\phi]$. In cases where $\mathrm{Det(M[\phi])}$ is positive useful 
algorithms can be found [1-3]. 
In other cases one can make progress only by using uncontrolled 
approximations \cite{Zha95}. Thus it is important to find alternative 
approaches to fermionic path integrals.

Recently a novel approach to solve certain lattice fermionic models 
was discovered [5-11].
It is possible to rewrite the partition function of these models as a sum 
over configurations of local bond variables with positive definite weights. 
The bonds $b$ connect lattice sites and thus divide the lattice into clusters. 
The partition function can be written as
\begin{equation}
Z \;=\; \sum_{[b]} W[b]\; \overline{\mathrm{Sign}}[b]
\end{equation}
where $W[b] > 0$ is the magnitude of the Boltzmann weight for
a given bond configuration, and $\overline{\mathrm{Sign}}[b] \geq 0$ is
an entropy factor arising from ``cluster flips'' that result due to
degrees of freedom other than the bonds''. This representation of
the fermionic partition function allows one to construct efficient cluster
algorithms which had previously been found only for bosonic problems.
In this article we describe this new approach to fermionic path 
integrals and present a result from recent Monte-Carlo studies using the 
new method.

\section{Fermion World-Line Path Integrals}
\label{fwl}

Consider spin-less fermions, hopping on a $d$-dimensional cubic lattice 
consisting of  $V = L^d$ sites and satisfying periodic or anti-periodic 
spatial boundary conditions. Let us focus on models whose Hamilton 
operators can be described by
\begin{equation}
H = \sum_{x,\hat{i}} h_{x,x+\hat{i}}
\label{ham}
\end{equation}
where $\hat{i}=1,2,..d$ represents directions and
$h_{x,i}$ is a nearest neighbor operator made up 
of the usual fermion creation and annihilation operators $c_x^+$ and $c_x$ 
associated with the site $x$. In order to write a path integral for such 
a problem, the Hamilton operator is decomposed into $2d$ terms 
\begin{equation}
H = H_1 + H_2 + ... + H_{2d},
\end{equation}
with
\begin{equation}
H_i = \!\! \sum_{\stackrel{x = (x_1,x_2,...,x_d)}{x_i\;even}} \!\! 
h_{x,x+\hat{i}} ~,~~
H_{i+d} = \!\! \sum_{\stackrel{x = (x_1,x_2,...,x_d)}{x_i\;odd}} \!\! 
h_{x,x+\hat{i}}.
\end{equation}
Note that the individual contributions to a given $H_i$ commute with each 
other, but two different $H_i$'s do not commute. Using the Suzuki-Trotter 
formula we can express the fermionic partition function as
\begin{eqnarray}
Z_f &=& {\rm Tr}\left[\mathrm{e}^{-H/T}\right] 
\nonumber \\
&\cong&
\sum_{n_1,n_2...} 
\langle n_1|\;[1-\epsilon H_1]\;|n_2\rangle\langle n_2|...
|n_{2d}\rangle \langle n_{2d}|\;[1-\epsilon H_{2d}]\;|n_{2d+1}\rangle
\nonumber \\
&& \langle n_{2d+1}|[1-\epsilon H_1]\;|n_{2d+2}\rangle\langle n_{2d+2}|
...|n_{2Md}\rangle\langle n_{2Md}|\;[1-\epsilon H_{2d}]\;|n_1\rangle
\end{eqnarray}
where the imaginary time extent $1/T$ has been divided into $M$ equal
steps of size $\epsilon = 1/T M$ and each of these steps have been
further divided into $2d$ time slices in each of which one of the 
$H_i$'s act individually. We have also used the complete set of
fermion occupation number states to evaluate the trace. 

The configuration of fermion occupation numbers $
[n] \equiv \{n_{x,t}\},t=1,2,...,2dM$ yields a fermion world line 
configuration. The path integral of the model is given by
\begin{equation}
Z_f \;=\; \sum_{[n]}\; W[n]\;\Sign[n]
\end{equation}
\nonumber \\
where the magnitude of the Boltzmann weight $W[n]$ is the product of 
the magnitude of transfer matrix elements and the
sign of the Boltzmann weight $\Sign[n]$ is the product of their signs. 
Practically, if we ignore the anti-commutation relations between fermionic
operators on different sites, the magnitude of the transfer matrix element
and hence $W[n]$ does not change. On the other hand $\Sign[n]$ 
turns into a product of signs of the new transfer matrix elements 
times a global sign factor that takes into account the signs 
arising due to anti-commutation relations that were ignored in
while calculating the transfer matrix elements.
This global sign factor is topological in origin and can
be found by tracking the permutation of the fermion world lines 
in time when the fermions are conserved \cite{Wie93}. The sign is 
positive for an even permutation and negative for an odd permutation.

\section{Meron Cluster Approach}
\label{mca}

The world-line approach to fermionic path-integral cannot be used
to design algorithms because the Boltzmann weight is not positive
definite and the correct probability density to generate world-line
configurations is not known. One typically needs an 
exponentially large amount of statistics to evaluate an expectation value 
using Monte-Carlo techniques. This is referred to as the sign 
problem. For example it is easy to check that the naive approach to 
evaluate
\begin{equation}
\langle O \rangle = \frac{1}{Z_f} \sum_{[n]} O[n]\; \Sign[n]\; W[n]
\end{equation}
where the partition function $Z_f$ is given by
\begin{equation}
Z_f = \sum_{[n]} \Sign[n]\;W[n]
\end{equation}
such that $W[n] > 0$ and $\Sign[n] \;=\;\pm 1$ suffers from a sign
problem. Fortunately, in most physically interesting problems there is 
a lot of freedom to choose the variables in which to express the
partition function and evaluate the observables. Using this freedom
some times one can be clever and find variables in which the
Boltzmann weights turn out to be positive so that the sign problem
is solved. 

Recently, non-local cluster variables have been successful 
is solving fermionic sign problems. The world-line partition function 
is first rewritten by introducing new ``bond'' variables that connect 
lattice sites in addition to the fermionic occupation variables that 
live on sites. Mathematically this means
\begin{equation}
Z_f = \sum_{[n]} \;\Sign[n]\;W[n]\;=\; \sum_{[n,b]}\;\Sign[n,b]\;W[n,b]
\end{equation}
where $[n,b]$ refers to the new configuration of fermions and bonds.
Bonds connect lattice sites into clusters and each bond configuration 
naturally divides all sites of the lattice into a collection of clusters.
New configurations can be obtained by reversing the fermion occupation on 
the sites associated with a single cluster. This is referred to as a 
{\em cluster flip}.

Clearly, there is a lot of freedom in choosing $\Sign[n,b]$ and
$W[n,b]$ such that the partition function remains unchanged. However,
if we restrict the choices such that
\begin{itemize}
\item[(a)] Cluster flips do not change $W[n,b]$, i.e., $W[n,b]\equiv W[b]$,
\item[(b)] Cluster flips effect $\Sign[n,b]$ independently,
\item[(c)] Cluster flips can always produce a positive configuration,
\end{itemize}
then we can ensure a complete solution to the sign problem since
it is possible to perform an average over cluster 
flips which leads to 
\begin{equation}
Z_f \;=\; \sum_{[b]}\overline{\Sign}[b]\;W[b],
\label{clpf}
\end{equation}
where 
\begin{equation}
\sum_{\rm{cluster flips}} \Sign[n,b]\;=\;\overline{\Sign}[b] = 0, 
2^{N_{\cal C}}
\label{clsign}
\end{equation}
and $N_{\cal C}$ is the number of clusters. 

A cluster whose flip changes the sign of a configuration is called
a {\em meron}. Properties (a) and (b) imply that such clusters identify
two configurations of equal weight and opposite signs and hence
do not contribute to the partition function. On the other hand, meron
clusters can contribute to observables. For example typically condensates 
get contribution from one meron sector, two point functions get 
contribution from zero, one and two meron sectors etc. An
algorithm for such a problem must generate bond configurations with
weight $W[b]$ but suppress meron clusters.

\section{Spin-less Fermions}

\begin{figure}[tb]
\centerline{\psfig{file=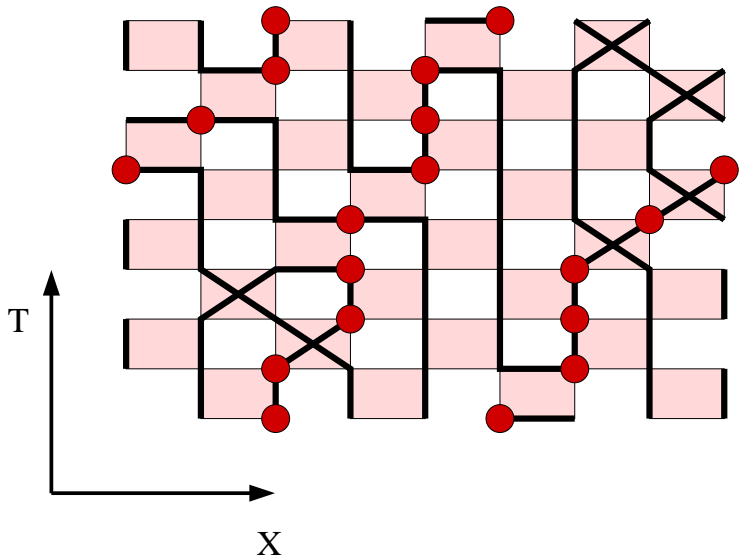,width=100mm}}
\caption{ A typical configuration of fermions and bonds.}
\label{fwlbconf}
\vskip-0.15in
\end{figure}

Let us illustrate the ideas of the previous section using a simple
example. Consider the nearest neighbor Hamilton
operator introduced in section \ref{fwl} with
\begin{equation}
h_{x,i} =  \left[ 
- t (c^+_x c_{x+\hat{i}} + c^+_{x+\hat{i}} c_x) + \mu (n_x + n_{x+\hat{i}})
+ U (n_x - \half)(n_{x+\hat{i}} - \half)\right].
\label{slhop}
\end{equation}
with the constraint $t = U/2 + \mu$. Here $n_x = c^\dagger_x c_x$, is 
the fermion number operator. 
\begin{figure}[b]
\centerline{\psfig{file=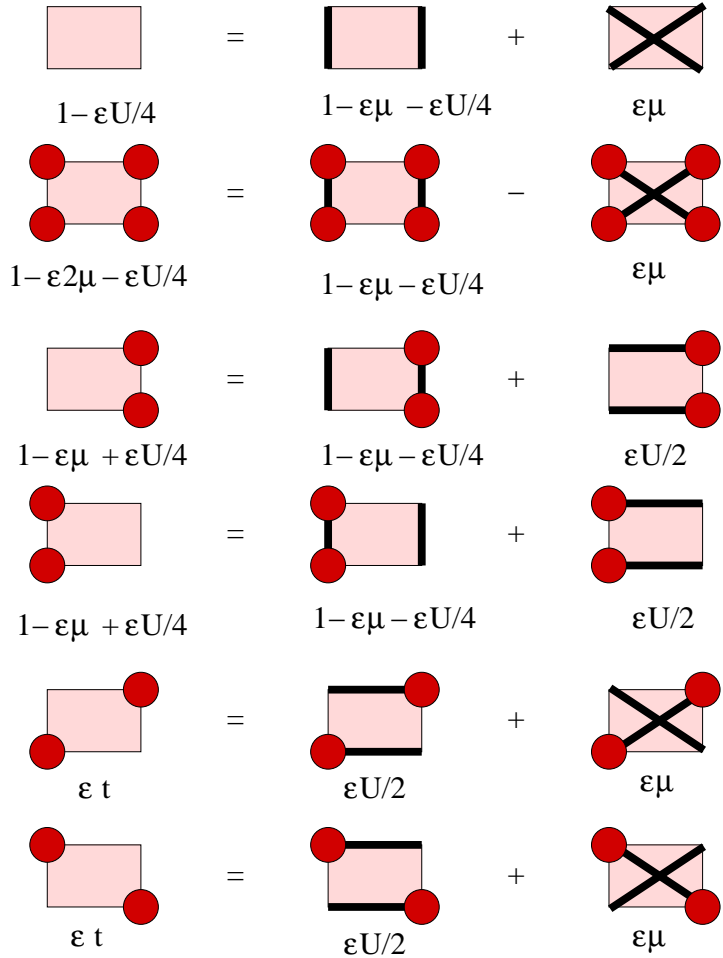,width=95mm}}
\caption{Interaction plaquettes and their weights that contribute to $W[n]$ 
and $W[n,b]$ so that $Z_f$ is preserved. It is assumed that $t = U/2 + \mu$.}
\label{breakup}
\end{figure}
Figure \ref{breakup} shows the non-zero transfer matrix elements 
$[1-\varepsilon h_{x,y}]$ in the occupation number configuration $[n]$
as well as in the extended configuration of occupation numbers and bonds
$[n,b]$, such that partition function does not change.
A typical configuration of fermion occupation numbers and bonds 
in one spatial and one temporal direction with $M=4$ is shown in Fig. 
\ref{fwlbconf}. The shaded regions represent the interaction plaquettes 
each of which represents a transfer matrix element. The global fermion 
permutation sign factor for this configuration is $-1$.

The significance of the breakup chosen in Fig. \ref{breakup} is that it 
yields a model that satisfies properties (a) and (b) of the meron cluster 
approach. The first property is easily checked by noticing that the 
magnitude of the weights of the transfer matrix elements depend only on 
the bonds and not on the fermion occupation numbers. In order to check that
the property (b) is also satisfied it is important to find the effect 
of a cluster flip on $\Sign[n,b]$. It can be shown that if
\begin{equation}
\frac{N_{\mbox{side hops}}}{2} + 
N_{\mbox{cross hop over filled sites}} + 
N_{\mbox{temporal winding}}
\label{meron}
\end{equation}
for a cluster is even only then the fermion permutation sign changes when 
that cluster is flipped. Details of the derivation of this formula can
be found in \cite{Cha00.2}. The formula shows that the cross bonds can 
in principle induce dependence between clusters. However, the extra local
negative sign associated with the fully occupied cross bond in 
Fig. \ref{breakup}, cancels such dependences. Finally for property (c) to 
be satisfied one has to either set $\mu = 0$ or $U=0$. 
When $\mu=0$ there are no cross bonds and one can flip all configurations 
$[n,b]$ to a configuration of staggered spatial fermion occupation which 
is static in time. This configuration is again guaranteed to be positive
\cite{Cha99.1}.
On the other hand when $U=0$ there are no horizontal bonds and one can flip 
all the configurations $[n,b]$ to a configuration with all sites are empty.
This configuration is guaranteed to be positive \cite{Cha99.2}. 

\section{Fermions with Spin}

\begin{figure}[tb]
\centerline{\psfig{file=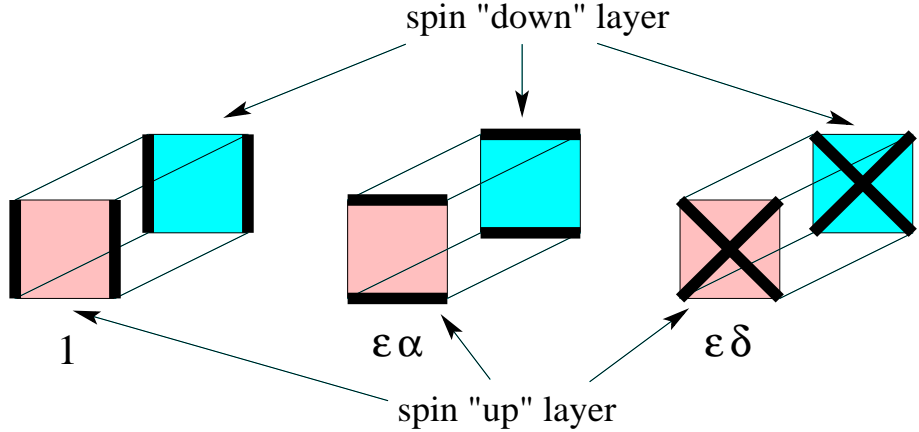,width=122mm}}
\caption{ Bond configurations and their weights of an interaction cube
for a model with spin.}
\label{intplaq}
\end{figure}

Fermionic Hamiltonians that are of interest in both nuclear and
condensed matter systems, involve internal degrees of freedom like
spin and isospin. In order to illustrate how the meron cluster approach
can be extended to such systems let us construct a model of fermions with 
spin which can be viewed as two layers of spin-less fermions.
Instead of starting from the Hamilton operator and then constructing
the cluster model as we did in the case of spin-less fermions, let us
construct the cluster model first. For simplicity, consider models where 
the bond configuration of Fig. \ref{fwlbconf} represents a typical 
configuration on each of the layers. Let the interactions between
layers arise due to a constraints on how the bonds in the two layers
are related. Fig. \ref{intplaq} shows the allowed 
bond configurations on each ``interaction cube'' of a simple interacting 
model and the magnitude of their weights.
This means that every configuration is forced to contain identical clusters 
in the two layers. In addition, if the allowed fermion occupation numbers 
on the sites for a given bond configuration satisfy the rules of the 
previous section (i.e., the occupation numbers of sites connected by 
vertical and cross bonds must be the same and those connected by 
horizontal bonds must be opposite) and if all these configurations 
have the same weight $W[b]$ ( which is the product of the
magnitudes of the interaction cube weights specified in Fig. \ref{intplaq})
then the model automatically satisfies the property (a) of section 
\ref{mca}. Since the clusters live on a single layer we can
again use eq. (\ref{meron}) to determine the sign change due to a cluster
flip. In order for the cross bonds not to violate property (b), we again 
associate an extra local negative sign with fully occupied cross bonds on 
each spin layer. It is easy to check that these restrictions 
automatically also satisfy property (c) and the cluster model
thus becomes solvable with the meron cluster approach.

As before the partition function of the model is given by eq. (\ref{clpf}) 
with $\overline{\Sign}[b]$ given by eq. (\ref{clsign}).
In the present case, the number of clusters $N_{\cal C}$ is always 
even with $N_{\cal C}/2$ clusters in the spin up layer and an equal number
of identical clusters in the spin down layer. When the quantity of 
eq. (\ref{meron}) is even for any of the clusters then 
$\overline{\Sign}[b] = 0$. Such clusters are the meron clusters. 
When there are no meron clusters in the configuration then
$\overline{\Sign}[b] = 2^{N_{\cal C}}$ since each cluster can have 
two fermion occupation number configurations associated to it. 
Retracing the steps of the previous section in the reverse order,
we can also determine the associated Hamilton operator. Remembering that 
$2dM$ is the number of time slices of the lattice and $\varepsilon M = 1/T$,
it can be shown that the model obtained in the limit 
$\varepsilon \rightarrow 0$ can be described by the Hamilton operator 
of eq. ({\ref{ham}) with
\begin{eqnarray}
\label{ham1}
&& h_{x,y} \;=\; 
-\delta
\left\{[{\cxd}^\dagger \cyd + {\cyd}^\dagger \cxd - \nxd - \nyd + 1]\;\;\;\;
\times \right.
\nonumber \\
&& \left.\;\;\;\;\;\;\;\;\;\;\;\;\;\;\;\;\;\;\;\;\;\;\;\;\;
[{\cxu}^\dagger \cyu + {\cyu}^\dagger \cxu - \nxu - \nyu + 1]
\right\}
\nonumber \\
&&-\alpha\left\{[{\cxd}^\dagger \cyd + {\cyd}^\dagger \cxd - 
2(\nxd - 1/2)(\nyd-1/2) + 1/2]\;\;\;\;\times\right. 
\nonumber \\
&& \left.\;\;\;\;\;\;\;\;\;[{\cxu}^\dagger \cyu + {\cyu}^\dagger \cxu - 
2(\nxu - 1/2)(\nyu-1/2) + 1/2]\right\}.
\label{rawh}
\end{eqnarray}
Here the spin up and down fermions are created and annihilated 
by $\cxu^+,\cxd^+$ and $\cxu,\cxd$ respectively and $\nxu$ and $\nxd$
refer to the corresponding number operators.

\begin{figure}[ht]
\vskip0.25in
\centerline{\psfig{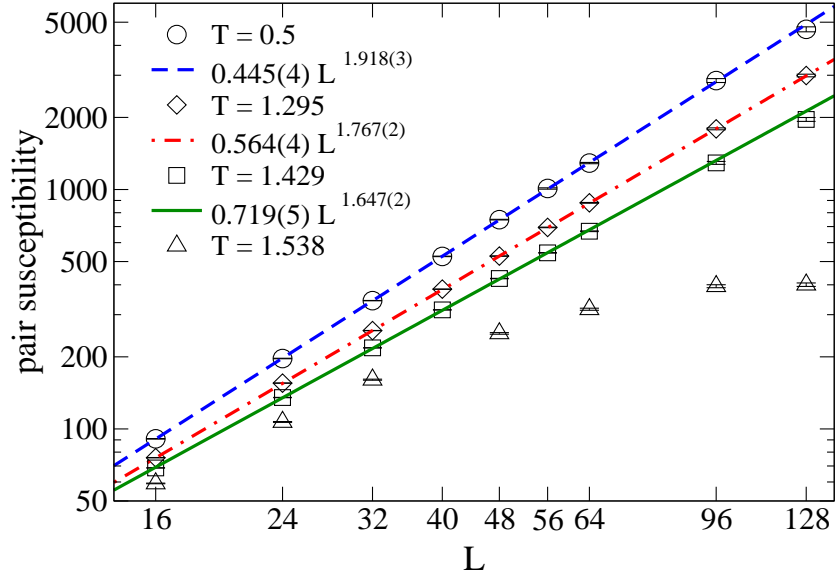}}
\caption{ Pair susceptibility as a function of $L$.}
\label{psus}
\end{figure}

The above Hubbard type model has a $U(1)$ fermion number symmetry. In 
three or more dimensions the breaking of this symmetry leads to 
superfluidity(or superconductivity when the symmetry is gauged). In 
two dimensions due to the Mermin Wagner theorem the 
symmetry cannot break spontaneously and the superfluid transition is 
driven by the Kosterlitz-Thouless phenomena \cite{Kos73}. 
It has been difficult to study the predictions of universality in
fermionic systems. This is the reason no known calculation exists
that confirms the predictions of the Kosterlitz-Thouless phenomena 
starting from a microscopic fermionic Hamiltonian. The above model
for the first time provides an opportunity for studying such a 
superfluid transition using cluster algorithms. The simplest 
observable relevant to this transition is the pair susceptibility which 
can be defined as
\begin{equation}
\chi \; = \;
\frac{2 T}{Z V} \; \int_0^{1/T} \;dt\;
 {\rm Tr} \left[ \; {\rm e}^{-(1/T-t) H}\; p^+ \;
 {\rm e}^{-t H}\; p^- \;\right]
\end{equation}
with $p^+ = \sum_x c_{x,\uparrow}^\dagger c_{x,\downarrow}^\dagger$
the pair creation and $p^- = (p^+)^\dagger$ the pair annihilation operators.
In terms of cluster variables the susceptibility is proportional to 
the sum over the square of the size of certain clusters depending on the 
number of meron clusters in the configuration. The Kosterlitz-Thouless 
prediction says that
\begin{equation}
\chi \;\propto \; \left\{\begin{array}{cc} L^{2-\eta(T)} & T < T_c \cr
					\mbox{Const.} & T > T_c
		\end{array}\right.
\label{fsspl}
\end{equation}
where $\eta(T)$ changes continuously from $0.25$ at $T_c$ to
$0$ at $T=0$. Figure \ref{psus} shows the results for the pair susceptibility 
as a function of spatial size $L$ in the above model with $\delta=1,\alpha=1,
d=2$ and $M=20$. As can be seen, $T=1.538,1.429$ are both above $T_c$, although
in the latter case lattices of size $L=128$ are necessary to see the 
saturation. At $T=1.295$ and $T=0.5$ the power law fits are extremely good 
over the entire range of $L$ with powers $1.767(2)$ and $1.918(3)$. 
More details can be found in \cite{Cha01.a}.

\section{Model Extensions}

Although the meron cluster approach helps to find models without sign 
problems, it is also possible to find cluster models which 
seem not to satisfy all the properties of the meron cluster approach 
but still do not suffer from sign problems. For example the model with 
the Hamilton operator 
$H = \sum_{x,i} h_{x,x+i} + \sum_x h_x$, where
\begin{equation}
h_x \;=\; U \left(\cxu - \frac{1}{2}\right)\left(\cxd - \frac{1}{2}\right)
+ \mu (\nxu + \nxd).
\end{equation}
will have the same partition function as eq. (\ref{clpf}) except that
$\overline{\Sign}[b]$ is no longer $2^{N_{\cal C}}$ but is replaced by
\begin{equation}
\prod_{{\cal C} \;\in\; \uparrow \;\mbox{layer}}
\left[
\mathrm{e}^{
(\epsilon/2d)\;\left(-U\;S_{\cal C}/4 + \mu \Omega_{\cal C}\right)}
+
\mathrm{e}^{
(\epsilon/2d)\;\left(-U\;S_{\cal C}/4 - \mu \Omega_{\cal C}\right)}\right]
\;\pm\;
2\;\mathrm{e}^{(\epsilon/2d)\;U\;S_{\cal C}/4},
\label{sfactor}
\end{equation}
where $S_{{\cal C}}$ is the size of the cluster ${\cal C}$ and 
$\Omega_{\cal C}$ is $2dM$ times the cluster's
temporal winding number. The negative sign should be taken for a 
meron cluster. This factor is obviously positive for any $\mu$ if $U < 0$.
Interestingly, it is also positive for $\mu < U/2$ if 
$\delta = 0$ and $U > 0$ since then meron clusters always come in pairs.
The proof also uses the fact that for all clusters 
$S_{\cal C} \geq \Omega_{\cal C}$. Thus we have found a 
repulsive Hubbard type model which does not suffer from a sign problem 
when formulated in the cluster approach for at least a limited range 
of chemical potentials. 

\section{Conclusions}

We have sketched how a variety of fermionic partition functions can
be written as a sum over cluster configurations with positive definite
weights. The three properties of the meron cluster approach discussed
in section \ref{mca} help find such models. However, it is possible to relax 
these properties and find extensions in certain cases. One drawback of 
this approach is that
the new models possess a complicated Hamilton operator. On the other hand, 
given the numerical efficiency of the algorithms that can be constructed 
for them it may still be useful to study these models. For example this 
approach has led to the first precise confirmation of universality 
arguments in fermionic systems [7-10].
Other applications in condensed matter and nuclear physics are being 
explored.

\section{Acknowledgment} 

I would like to thank J. C. Osborn and U.-J. Wiese for their collaboration.
This work is supported in part by funds provided by US Department of 
Energy grant DE-FG02-96ER40945 and the National Science Foundation
grant DMR-0103003.


\end{document}